# Free Volume Power Law for Transport Properties of Hard Sphere Fluid

**Hongqin Liu***

Integrated High Performance Computing Branch, Shared Services Canada, Montreal, QC, Canada

**Abstract**

This paper presents a study on the relationship between transport properties and geometric free volume for hard sphere (HS) system in dense fluid region. Firstly, a generic free volume distribution function is proposed based on recent simulation results for the HS geometric free volume by Maiti et al.[1,2] Combining the new distribution function with a local particle transportation model, we obtain a power law for the HS transport properties. Then a relation between the geometric free volume and thermodynamic free volume is established, which makes it possible to use well-developed equations of state (EoS) for the expressions of the geometric free volume. The new power law models are tested with molecular dynamic (MD) simulation results for HS viscosity, diffusivity and thermal conductivity, respectively and the results are very satisfactory. Using the power law we are able to reproduce several equations obtained from different approaches, such as the entropy scaling laws[3], mode coupling theory[4] or empirical correlations[5]. In particular, A long-standing controversy regarding the well known Cohen-Turnbull-Doolittle free volume model[6,7] is resolved by using the power law combined with an EoS.

* Email: hongqin.liu@canada.ca; hqliu2000@gmail.com

## 1. Introduction

Hard sphere (HS) fluid plays a central role in the study of condensed matter properties. Transport properties in dense fluids, especially liquids, i.e. viscosity (or fluidity), diffusion coefficient, and thermal conductivity are dominated by repulsive interactions[8,9], therefore, accurate descriptions of HS fluid set a solid foundation for real fluids. Based on HS equations for diffusion coefficient and viscosity, models for the Lennard-Jones (LJ) fluid and real fluids can be developed, typically, by using a so-called effective diameter method, where the HS diameter is "softened" with a temperature- (and density-) dependent expression for the repulsive contribution[8-10].

Due to the complexity of the problem, physically sound solutions only work for dilute or intermediate dense gases. For liquids, semi-empirical approaches or empirical correlations are mostly employed[5,9]. An example is the Cohen-Turnbull (CT) free volume theory[6], which tries to set up a physical foundation for the free volume exponential law proposed empirically by Doolittle[7]. It is one of the most employed models for liquids, supercooled liquids and glasses[9,11,12]. However, there exist some fundamental drawbacks and controversy regarding the theory. (1) The free volume is not well defined[13]. In the model[6,7], the free volume is a van der Waals (vdW) type. The vdW fee volume is only applicable to low density gas while the final expression works for liquid phase. (2) The most important contribution of the CT theory is arguably the free volume distribution function obtained by a statistical method[6]. However, the free volume used in the distribution function was not defined either and the function itself lacks solid justifications. (3) As pointed out by Goldstein[13], assumptions used to derive the theory ignored the relaxation process and therefore suffers a basic shortcoming.

In this work, a well-defined quantity, the geometric free volume[1,2,14,15] is adopted and we show that transport properties of dense HS fluids are related to this quantity. Throughout this paper, "dense fluid" is defined as high dense gas and, in particular, liquid. For developing new model, a reliable free volume distribution function is required. A great amount of computer simulation data for the distribution function have been reported in last decade or so[,1,2,15], which paves a way for deriving a physically sound distribution function. With this new distribution function and a local particle transportation model that takes into consideration of Goldstein's suggestions[13], we derive a free volume power law. Extensive testing and discussions are carried out for the new model. We also show that the free volume exponential law is not suitable for the transport properties as appropriate free volume expressions are employed. For self-completeness, we start with a short review since some equations will be involved.

## 2. A short review of related free volume models

In this section, the term "free volume" refers to the conventional concept, which has not been strictly defined in general and depends on the circumstances or the context. Mostly it refers to the van der Waals type free volume, which works only for low density gases, $V_f = V - V_0$. The simplest free volume model is the empirical equation proposed by Batchinski[16] and modified by Hildebrand[17], which reads:

$$\frac{1}{\mu} = \emptyset \ (or \ D) = AV_f = B\frac{V - V_0}{V_0} \quad (1)$$

where $\mu$ is viscosity, $\emptyset$, known as fluidity, $V_f$, the molar free volume, $A$ is a weak function of temperature: $A \sim T^{1/2}$, $B$ and $V_0$ are parameters. In this paper, upper case is used for molar volumes, lower case for molecule/particle level: $v = V/N_a = 1/\rho$, or $v_f = V_f/N_a$, where $N_a$ is the Avogadro constant. Ertl and Dullien (1973)[18] empirically modified Eq.(1), and propose a power law for diffusion coefficient:

$$D = A_0 V_f^m = B\left(\frac{V - V_0}{V_0}\right)^m \quad (2)$$

Where $A_0$ and $m$ are parameters. The third parameter, $m$, will definitely improve the correlation. Doolittle (1951)[7] empirically proposed the well know free volume model for viscosity (μ) and diffusion coefficient (D):

$$\frac{1}{\mu} = \emptyset = A \exp\left(-\frac{B_0}{V_f}\right) \quad (3)$$

where $A$ and $B_0$ are parameters, or $A \sim T^{1/2}$. In all the above empirical correlations, the vdW-type free volume is adopted:

$$V_f = V - V_0, or \ \frac{V_f}{V} = \frac{v_f}{v} = 1 - c\eta \quad (4)$$

where $\eta = \frac{1}{6}\pi\rho\sigma^3 = \frac{1}{6}\pi\rho^*$, $V_0$ (or $c$) is an adjustable parameter for real fluids, $\sigma$, the diameter of the particle, and for the van der Waals fluid, $c = 4$. It should be mentioned that Eq.(1)-(3) are proposed for real fluids, not limited to the HS fluid. Eq.(3) has been very successful when three or even more[11] parameters are employed. Use of empirical parameters will improve correlations, but physical meaning may be overlooked or undermined. The Cohen-Turnbull- Doolittle model[6,7] is a typical example.

Due to the successes achieved by the Doolittle's equation[7], Cohen and Turnbull (CT) (1959)[6] tried to derive the equation based on some theoretical arguments. Firstly, using a statistical method, they derived a free volume distribution function:

$$f(v_f) = \left(\frac{\gamma}{v_f}\right) exp\left(\frac{-\gamma v_l}{v_f}\right) \quad (5)$$

where $v_l$ is defined as the local free volume, which depends on particle position or local density; $v_f$, the average free volume, which is a function of the mean density, $\rho$. The free volume in the CT theory was not explicitly defined, but the final equation, Eq.(3), implicates that it is Eq.(4). Nevertheless, this was the first effort to derive a free volume distribution function and, as shown later, it does provide reasonable predictions as appropriate free volume is employed. Due to lack of reliable measures at the time, Eq.(5) could not be properly justified or tested. The authors then suggested a simple model for the local diffusivity (or fluidity) on the local free volume, using a threshold, $v^*$:

$$D(v_l) \sim \begin{cases} v_l, & v_l \geq v^* \\ 0, & otherwise \end{cases} \quad (6)$$

Finally,

$$D = \int_{v^*}^{\infty} D(v_l) f(v_l) dv_l = A\, exp\left(\frac{-\gamma v^*}{v_f}\right) \quad (7)$$

where $\gamma$ is a parameter, the so-called overlapping factor, and $A \sim \sqrt{T}$. The equation can be rewritten as:

$$\frac{D}{D_0} = exp\left(\frac{-\gamma v^*}{v_f}\right) = exp\left(\frac{-\gamma_0 \rho^*}{v_f/v}\right) \quad (8)$$

where, $D_0$ is the diffusivity for dilute gas, $\propto \sqrt{T}$, $\rho^* = \rho\sigma^3 = \sigma^3/v$. A similar equation can be written for viscosity (fluidity):

$$\frac{\emptyset}{\emptyset_0} = exp\left(\frac{-\gamma_1 \rho^*}{v_f/v}\right) \quad (9)$$

If the van der Waals free volume, Eq.(4), is adopted, the original Doolittle equation is recovered. For now we can assume that the CT theory, Eq.(8), (9), does not necessarily work only with the van der Waals free volume. From the derivation of the distribution function, Eq.(5), we may assume that the free volume $v_f/v$ could have a different definition since it aimed at liquid.

### 3. Free volume distribution function

Hereafter, by free volume, we mean the geometric free volume, which is defined[14,19] as the volume over which the centre of a given sphere can translate, given that the other N - 1 spheres are fixed. We use the same notation as the one used in previous section. But it should not be mixed up with each other.

The first effort to simulate geometric free volume distribution function was made by Sastry et al. (1989)[14] for the density range $\rho^*$=0.8, to 0.96. Debenedetti and Truskett (1999)[20] further discussed the issue and proposed a distribution function [14,20]:

$$f(v_l) \propto v_l^{\alpha'} exp\left(-\beta' v_l^{\gamma'}\right) \quad (10)$$

Where $\alpha'$, $\beta'$ and $\gamma'$ are *density-dependent* parameters. Krekelberg et al. (2006)[15] reported their simulation results for the free volume distribution function for a density range from 0.716 to 1.0. More recently, Mait et al.(2013,2014)[1,2] carried out simulations over a slightly wider density range: $\rho^*$=0.7, to 1.02. Their simulation results can be accurately represented with the following distribution function[2]:

$$f(v_l) = \frac{\gamma' \beta'^{\left(\frac{\alpha'+1}{\gamma'}\right)}}{\Gamma\left(\frac{\alpha'+1}{\gamma'}\right)} v_l^{\alpha'} exp\left(-\beta' v_l^{\gamma'}\right) \quad (11)$$

where $\Gamma(…)$ is the Gamma function. The authors[2] provided detailed values for $\alpha'$, $\beta'$ and $\gamma'$ at each density, from which the data used in this work are reproduced. The average free volume at each density, can be calculated by the following:

$$v_f = \int_0^{\infty} v_l f(v_l) dv_l \bigg/ \int_0^{\infty} f(v_l) dv_l \quad (12)$$

where $v_f$ and $v_l$ are in unit of $\sigma^3$. It is found that the simulation data sets from different sources (2014)[1], (2013)[2], (1998)[14], (1999)[20], (2006)[15] are all consistent with each other, and therefore, in this work our new distribution function will be built mainly on the data sets from ref [1,2].

Since the parameters, $\alpha'$, $\beta'$ and $\gamma'$, are density-dependent, eq.(11) or Eq.(10) cannot be used for our purpose. A new distribution function with parameters independent of density is required. Krekelberg et al. (2006)[15] noticed that the distribution function collapses into a single line when plotted against reduced free volume, $v_l/v_f$. By taking into account of the original function derived by Cohen and Turnbull[6], eq.(5), the free volume ratio, $v_l/v_f$, is used here, and combining it with eq.(11), the following new equation is proposed:

$$f\left(\frac{v_l}{v_f}\right) = \frac{b}{v_f}\left(\frac{v_l}{v_f}\right)^{\alpha} exp\left[-\beta\left(\frac{v_l}{v_f}\right)^{\gamma}\right] \quad (13)$$

where the parameters are found to be: $\alpha = 0.275$, $\beta = 3.18$, $\gamma = 0.47$ for the density range under consideration and from the normalization condition:

$$b = \gamma \beta^{\frac{\alpha+1}{\gamma}} / \Gamma\left(\frac{\alpha+1}{\gamma}\right) = 6.947 \quad (14)$$

All simulation results over the density range $\rho^* = 0.7 - 1.02$ can be reproduced remarkably well with the generic distribution function, Eq.(13). By the way we can also derive a similar distribution function for the HS surface area distribution using the simulation data[2,15]. The function takes the same form as Eq.(13) where $v_l/v_f$ is replaced by $v_{ls}/v_{fs}$ with the subscript "s" referring to the surface area, and the constants are given by $\alpha = 1.1981, \beta = 3.2736, \gamma = 0.71381$ and $b = \gamma \beta^{\frac{\alpha+1}{\gamma}} / \Gamma\left(\frac{\alpha+1}{\gamma}\right) = 12.7688$. Excellent predictions for the surface area distribution are also obtained.

**Figure 1** illustrates a comparison between the free volume distribution function reproduced by Eq.(13) and the raw data fitting[1,2], Eq.(11) with density-dependent parameters. Only at low density region, the differences are visible, otherwise invisible. By considering the fact that all three parameters in eq.(13) are density-independent, the agreement is remarkable.

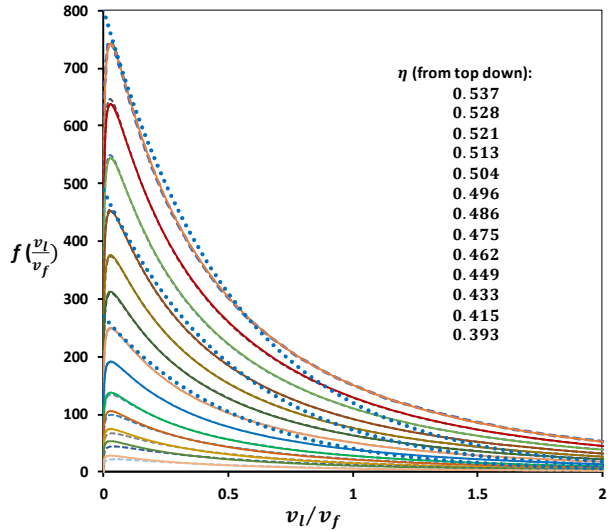

**Figure 1**. Free volume distribution function at different densities. Solid lines: calculated by Eq.(13). Dashed lines: calculated by eq.(11) with density-dependent parameters[2]. Dotted lines: CT distribution function, eq.(5), at $\eta = 0.537$, $\eta = 0.513$ and $\eta = 0.486$, where the function is adjusted to $\alpha_1 f(v_l)$, and two parameters are used to fit the data: $\alpha_1 = 1.55$, and $\gamma = 1.9$, without loosing the basic feature.

The prediction by the CT function, Eq.(5), is also illustrated. From Figure 1, the following observations are in order: (1) the CT function fails to predict the maximum values at $v_l/v_f \sim 0.027$, which is not a surprise since Eq.(5) is a monotonic function. (2) At high value range $v_l/v_f > 0.6$, the prediction is poor. (3) By recalling that in the original CT theory, the free volume was not defined. Here we plot it in the $v_l/v_f \sim f(v_l/v_f)$ plane and found out that it is not very bad at all. The last observation suggests that in the CT distribution function the undefined free volume can be considered as the geometric free volume.

With the generic distribution function, Eq.(13), we can perform some useful analysis. For example, to find the probability of all local free volume less than the average, $v_l \leq v_f$

$$P(v_l \leq v_f) = \int_0^{v_f} f(v_l)dv_l \bigg/ \int_0^\infty f(v_l)dv_l =$$
$$\frac{1}{\Gamma\left(\frac{\alpha+1}{\gamma}\right)} \int_0^1 t^{\left(\frac{\alpha+1}{\gamma}-1\right)} e^{-t}dt \approx 0.117 \quad (15)$$

Namely, around 11.7% of free volume falls into the range of $v_l \leq v_f$. At $v_l = v_f(\alpha/\beta\gamma)^{1/\gamma}$, or $v_l/v_f \approx 0.027$, the distribution function reaches a maximum.

Now we propose a procedure to obtain expressions for the geometric free volume. One could directly correlate the free volume with density by using some empirical functions. The risk is that empirical correlation may fail badly as the density falls beyond the correlation range. Here we establish a relation based on some physical arguments and the final correlation will be applied to the entire density range for consistency. From the generic distribution function, eq.(13), we define following entropic quantity:

$$F = \int_0^\infty f\left(\frac{v_l}{v_f}\right) \ln\left[f\left(\frac{v_l}{v_f}\right)\right] dv_l =$$
$$\ln\left(\frac{b}{v_f}\right) + \alpha b I_1 - b\beta I_2 \quad (16)$$

where

$$I_1 = \int_0^\infty x^\alpha \ln(x) e^{-\beta x^\gamma} dx =$$
$$\frac{1}{\gamma b}\left[\Psi\left(\frac{\alpha+1}{\gamma}\right) - \ln(\beta)\right] \quad (17a)$$

where $\Psi(.)$ is the digamma function, and

$$I_2 = \gamma^{-1}\beta^{-\frac{\alpha+\gamma+1}{\gamma}} \Gamma\left(\frac{\alpha+\gamma+1}{\gamma}\right) \quad (17b)$$

Obviously, both $I_1$ and $I_2$ are constants. For the applications discussed below, it is convenient to use a "reduced" form. At the dilute (ideal) gas limit: $\rho^* \to 0, v_f \to v, F \to const$. Therefore, we have:

$$F = ln\left(\frac{v_f}{v}\right) + const \quad (18)$$

In a study on distribution function for Voronoi free volume, Kumar and Kumaran (2005)[21] adopted the Grest and Cohen's information entropy expressed with the Voronoi-cell free volume distribution[11] and pointed out that at equilibrium the information entropy is identical to the thermodynamic entropy. Accordingly, they express the excess entropy in terms of the Voronoi-cell free volume distribution function[21]:

$$\frac{s^{ex}}{k} = -C \int_0^\infty f_c(v_{cf}) ln[f_c(v_{cf})] dv_{cf} \quad (19)$$

where $C$ is a constant, $v_{cf}$ is the local cell free volume. Here we conjecture that, in analog to Eq.(19) (see also ref 11), the excess entropy can also be expressed in terms of the geometric free volume:

$$\frac{s^{ex}}{k} = -C_0 \int_0^\infty f\left(\frac{v_l}{v_f}\right) ln\left[f\left(\frac{v_l}{v_f}\right)\right] dv_l \quad (20)$$

where $C_0$ is a constant. The thermodynamic free volume, $v_{tf}$, is defined in terms of the excess entropy:

$$ln\left(\frac{v_{tf}}{v}\right) = \frac{s^{ex}}{k} \quad (21)$$

Combining eq.(18), (20) and (21), we obtain a relation between the geometric free volume and thermodynamic free volume $v_{tf}$:

$$\frac{v_f}{v} = c_0 \left(\frac{v_{tf}}{v}\right)^\varsigma \quad (22)$$

where $c_0$ and $\varsigma$ are constants. This is one of the basic relations established in this work. Apparently eq.(22) is applicable to dense fluid region since the distribution function, Eq.(13), has been obtained for dense fluid. As shown below, the constant, $c_0$, will be "absorbed" into the coefficients in the final equations.

An obvious advantage of using Eq.(22) is that we have many well-established equations of state (EoS) for HS fluid from which the thermodynamic free volume can be obtained:

$$ln\frac{v_{tf}}{v} = -\int_0^\rho (Z-1)\frac{d\rho}{\rho} \quad (23)$$

where $Z$ is the compressibility. Recently current author proposed a new EoS for hard sphere fluid[22]:

$$Z = \frac{1 + \eta + \eta^2 - \frac{8}{13}\eta^3 - \eta^4 + \frac{1}{2}\eta^5}{(1-\eta)^3} \quad (24)$$

which is derived from a virial coefficient correlation as a modification to the Carnahan Starling (CS) equation[23]. Eq.(24) improves the accuracy by almost two-orders of magnitude compared to the CS EoS[23]. From Eq.(23) and (24) we have:

$$ln\frac{v_{tf}}{v} = \frac{5}{13}ln(1-\eta) - \frac{188\eta - 126\eta^2 - 13\eta^4}{52(1-\eta)^2} \quad (25)$$

Figure 2 depicts the testing results for Eq.(22), where $v_{tf}/v$ is calculated by Eq.(25). The values of the average free volume, $v_f/v$, are calculated for the local free volume data[2,14] by using Eq.(12). This figure can be read together with Figure 10b for a better understanding. The simulation data at low density end become less reliable and the point at $\rho^* = 0.7$ is discarded in the correlation (Figure 10b). The equation shown in the figure is from the Excel fitting. The observations are: (1) the simulation data from Ref [2] and Ref [14] are highly consistent with each other; (2) the linear prediction, $ln(v_{tf}/v)$ vs $ln(v_f/v)$, by Eq.(22) works satisfactory for the density range from $\rho^* \sim 0.7+$ to 1.02.

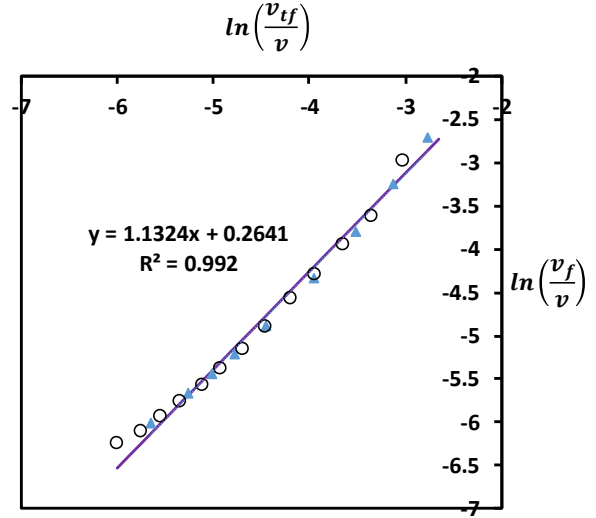

**Figure 2** Correlation between geometric free volume and thermodynamic free volume. Solid line: eq.(22); Circles: MD data by Sastry et al. (1989)[14]; Triangles: MD data by Maiti et al. (2013)[2].

Some other types of free volume have also been used[14,24-26], such as void space $v_s$ and available space, $v_a$. The relationships between them and the geometric free volume have been discussed[14,24-26]. In particular, the available space is strictly related to chemical potential[26], or excess entropy in the HS fluid. By using the similar arguments addressed above, we can propose a generic relation:

$$\left(\frac{v_s}{v}\right)^{c_1} \sim \left(\frac{v_a}{v}\right)^{c_2} \sim \left(\frac{v_f}{v}\right)^{c_3} \quad (26)$$

where $c_1, c_2$ and $c_3$ are constants. The most important point is that all these "free volumes" exhibit the similar trends or curvatures of density dependency in the density range we are interested.

### 4. A new free volume model

Since a reliable free volume distribution function, eq.(13), is now available from MD simulation, we are ready for developing a new model for the HS transport properties. To this end, we first quote Goldstein (1969)[17]: "For a hole or free volume mechanism to be complete, not only must the motion of molecules into holes be considered, but also the local appearance and disappearance of holes ... the molecular rearrangements involved in a molecular jump into a hole and those involved in the creation or vanishing of a hole are quite different, the latter seeming to require a much greater cooperative character". Accordingly, we propose a two-step transportation process. In the first step, a hard sphere finds a free volume with a minimum size of $v^*$, and jumps into it. In the second step, the free volume, $v^{\neq}$, left by the moving sphere, is relaxed. One possible way for the later to happen is that a neighbour sphere fills in this free volume. Figure 3 depicts this process.

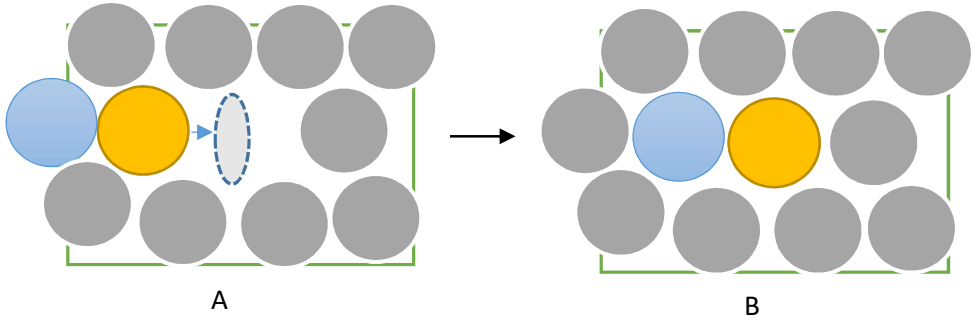

**Figure 3.** A local particle (yellow) transportation process. A: density fluctuation creates a local free volume with an area of order of $v_l^{2/3}$ (dashed); B: the local region becomes stable again after the 2-step diffusion is completed. Step (1): the yellow particle jumps into a free volume with a suitable area, $v_l^{2/3}$, $v_l \geq v^*$, dashed circle. Step (2): the free volume left by yellow particle is filled up by the light blue particle.

We use fluidity as an example. For step (1), with the same arguments used by Cohen and Turnbull for deriving Eq.(7)[6], we have:

$$\emptyset_1 = \int_{v^*}^{\infty} \varphi(v_l) f(l) dv_l \quad (27)$$

For step (2), it is calculated the same way as that in step (1), but the integration is from 0 to $v^{\neq}$, namely,

$$\emptyset_2 = \int_{0}^{v^{\neq}} \varphi(v_l) f(v_l) dv_l \quad (28)$$

which aligns with Goldman's comment since it has much less probability. It should be mentioned that $v^{\neq}$ may not be the same as $v^*$, but they should be close. Therefore, assuming $v^{\neq} \approx v^*$, we have the following:

$$\emptyset = \int_{v^*}^{\infty} \varphi(v_l) f(l) dv_l + \int_{0}^{v^{\neq}} \varphi(v_l) f(v_l) dv_l \approx \int_{0}^{\infty} \varphi(v_l) f(v_l) dv_l \quad (29)$$

The additivity of the two steps comes from the fact that retention time is additive. Eq.(29) effectively removes the intrinsic structural property of the fluid, $v^*$, used in the original CT model. This makes more sense since a macroscopic property should not explicitly depend on some intrinsic structural character.

Now we need to make another assumption: the dependence of local fluidity on free volume, namely the function $f(v_l)$. In the CT theory, a simple linear dependence, Eq.(6), is used. There are other options. In an effort to develop a new free volume model for diffusivity, Ricci et al. (1977)[27] assumed that the local diffusivity is proportional to the "free path": $D(v_l) \sim v_l^{1/3}$. As shown by Fig.3, when the yellow particle jumps into a free volume, it "sees" or "fits" the area, namely, the "free area": $v_l^{2/3}$, neither the whole free volume, nor the free path. Nonetheless, here we temporarily use a more generic form and leave the constant to be determined later:

$$\varphi(v_l) = c v_l^n \quad (30)$$

where $c$ and $n$ are constants. Eq.(13), (29) and (30) yield a Gamma function:

$$\emptyset = \frac{cb}{\gamma \beta^{\frac{\alpha+n+1}{\gamma}}} v_f^n \int_0^\infty t^{\frac{\alpha+n+1}{\gamma}-1} e^{-t} \frac{dt}{t} \quad (31)$$

where $t \equiv \beta(v_l/v_f)^{1/\gamma}$. Since $\gamma(= 0.47) < 1, \alpha(= 0.275) > 0, n > 0$, the Gamma function converges. The final result reads:

$$\emptyset = \frac{cb\Gamma\left(\frac{\alpha+n+1}{\gamma}\right)}{\gamma \beta^{\frac{\alpha+n+1}{\gamma}}} v_f^n \quad (32)$$

Eq.(32) is our final result. The density-independent coefficient can be eliminated by using a reduced form. At low density ($\rho^* \to 0$):

$$\emptyset \to \emptyset_0 \text{ as } v_f \to v \quad (33)$$

where the viscosity of dilute gas is given by[5]:

$$\mu_0 = \frac{1}{\emptyset_0} = \frac{5}{16\sigma^2}\left(\frac{mkT}{\pi}\right)^{\frac{1}{2}} = \frac{5}{16\pi^{\frac{1}{2}}} \frac{(m\varepsilon T^*)^{\frac{1}{2}}}{\sigma^2} \quad (34)$$

where $m$ is the mass, $k$, the Boltzmann constant, T, temperature, $\epsilon$, the energy parameter, and $T^* = kT/\epsilon$. Eq.(34) is given here since it is required in converting the MD data for both HS and LJ fluids. By the way, the Enskog theory for viscosity reads[5,9]:

$$\frac{\mu_E}{\mu_0} = \frac{1.016}{g(\sigma)} + 0.8\left(\frac{2\pi}{3}\rho^*\right) + 0.7737 g(\sigma)\left(\frac{2\pi}{3}\rho^*\right)^2 \quad (35)$$

where $g(\sigma)$ is the radial distribution function at contact. Eq.(35) is required when converting the data from Alder et al.(1970)[28]. Finally, the reduced fluidity is obtained from eq.(32) and (33):

$$\emptyset^* \equiv \frac{\emptyset}{\emptyset_0} = \left(\frac{v_f}{v}\right)^n \quad (36)$$

Now we come back to the constant $n$ in Eq.(36). With Cohen and Turnbull's assumption, Eq.(6), $n = 1$. For fluidity,

$$\frac{\emptyset}{\emptyset_0} = \frac{v_f}{v} = \frac{V_f}{V} \quad (37)$$

which is the Hildebrand-Batstinski model, Eq.(1) if the vdW free volume is applied (inappropriately). For hard sphere fluid, as mentioned (Figure 3), the particle only "sees" the area, namely, $n = 2/3$:

$$\frac{\emptyset}{\emptyset_0} = \left(\frac{v_f}{v}\right)^{\frac{2}{3}} \quad (38)$$

The case with diffusion coefficient is different: both $D/D_E$ and $D/D_0$ are not simple monotonic functions of density. $D/D_E$ exhibits a maximum at $\rho^* \sim 0.5$[8] and $D/D_0$ is a slightly "bow-shaped" curve (see Fig.4).

Therefore, we need to seek some other reduced form if a simple free volume power law can also be established for diffusion coefficient.

In searching for a scaling law for diffusion coefficient with entropy, Dzugutov (1996)[29] introduced a dimensionless form by using the Enskog theory for the collision frequency, $\Gamma_E = 4\sigma^2 g(\sigma)\rho(\pi kT/m)^{1/2}$. Following the same arguments, we use the following dimensionless form for diffusion coefficient:

$$D^* \equiv \frac{D}{D_0 g(\sigma)} \quad (39)$$

Then the free volume model for hard sphere diffusivity reads

$$D^* = \left(\frac{v_f}{v}\right)^\kappa \quad (40)$$

Due to introducing $g(\sigma)$, we leave the constant $\kappa$ to be determined by fitting MD data and it is found that $\kappa = 0.74$ for the HS fluid. The radial distribution function at contact $g(\sigma)$ is calculated with a HS EoS and here we use Eq.(24):

$$g(\sigma) = \frac{Z-1}{4\eta} = \frac{1 - \frac{1}{2}\eta + \frac{5}{52}\eta^2 - \frac{1}{4}\eta^3 + \frac{1}{8}\eta^4}{(1-\eta)^3} \quad (41)$$

The diffusion coefficient for dilute gas (for data conversion) is given by[5]:

$$D_0 = \frac{3}{8\rho\sigma^2}\left(\frac{kT}{\pi m}\right)^{\frac{1}{2}} = \frac{3}{8\pi^{1/2}\rho^*}\left(\frac{T^*\sigma\epsilon}{m}\right)^{\frac{1}{2}} \quad (42)$$

Eqs.(36), (38) and (40) are the power-law free volume models for transport properties proposed in this work. For the HS fluids, $n = 2/3, \kappa = 0.74$ for viscosity, and diffusivity, respectively. With the thermodynamic free volume, our new models can be written as:

$$\frac{\emptyset}{\emptyset_0} = \left(\frac{v_f}{v}\right)^{\frac{2}{3}} = \left(\frac{v_{tf}}{v}\right)^\xi \quad (43)$$

where the constant is obtained from $\xi = 1.132(2/3) \approx 0.75$ (Figure 2). This is a remarkable result since it is entirely predictive. For diffusion coefficient we have:

$$D^* = \left(\frac{v_f}{v}\right)^{0.74} = \left(\frac{v_{tf}}{v}\right)^\chi \quad (44)$$

where the best-fitted value: $\chi = 0.836$ for the HS fluid.

Lastly, despite the fact that there is not much physical justifications for thermal conductivity to be treated the same way as diffusion coefficient and viscosity, there are some researchers who adapt a similar model for thermal conductivity (see Sigurgeirsson and Heyes, 2003[5] and a review[9]). Here we propose the following power law for HS thermal conductivity.

$$\lambda^{*-1} \equiv \frac{(\lambda/\lambda_0)^{-1}}{g(\sigma)} = \left(\frac{v_{tf}}{v}\right)^{\delta} \quad (45)$$

where

$$\lambda_0 = \frac{75}{64\sigma^2}\left(\frac{kT}{\pi m}\right)^{\frac{1}{2}} \quad (46)$$

For the MD simulation data conversion, the Enskog equation is also required:

$$\frac{\lambda_E}{\lambda_0} = \frac{1.025}{g(\sigma)} + 1.23\left(\frac{2\pi}{3}\rho^*\right) + 0.776 g(\sigma)\left(\frac{2\pi}{3}\rho^*\right)^2 \quad (47)$$

Eq.(45) is only an empirical analog to the diffusion coefficient model. The parameter in eq.(45) is obtained from fitting the MD data, $\delta = 1.2$.

## 5. Testing with MD data for HS and LJ fluids

Now we revisit the CT model, Eq.(8) and Eq.(9). Since the vdW-type free volume, Eq.(4), is not applicable to dense fluid, we use the geometric free volume for testing the CT theory, which is justified by Figure 1. For a strict testing, there is only one parameter, $\gamma_1$, in Eq.(9) that can be adjusted and the expression, Eq.(22) with (25), is given for the HS free volume and the parameters within, if any, should not be altered by transport properties. Figure 4 depicts the results where two solid lines are for diffusion coefficient and fluidity, respectively. The dashed lines show how the CT model prediction, Eq.(9), changes as the parameter $\gamma_1$ is altered. An immediate observation is that there is no way the CT theory can represent the HS transport properties. In other words, the free volume exponential law is not compatible with the change of the HS transport properties. Then why it works in practical applications? We will come to this later.

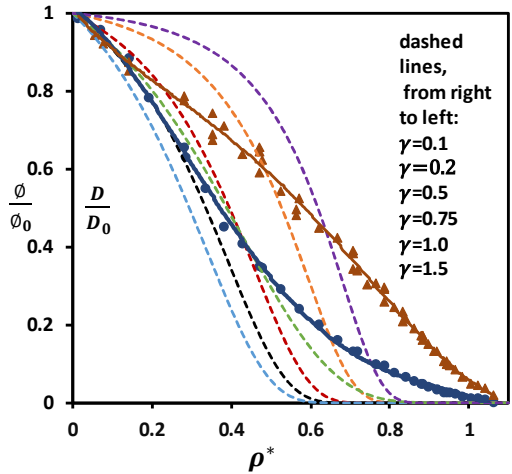

**Figure 4**. Incompatibility of the CTD model, eq.(9), with the simulation data. The dashed lines are calculated by Eq.(9). The MD simulation results (see Fig.5 and Fig.6) are fitted with 4$^{th}$ order polynomial functions (solid lines). The parameter, $\gamma$, represents the parameter in Eq.(8) or (9).

For testing the new models, we need simulation data for HS transport properties. Ever since the pioneer work of Alder et al.(1970)[28], numerous MD simulation results have been reported. Most of them are consistent with each other. In particular, Sigurgeisson and Heyes (2003)[5] reported their high quality MD results for HS diffusion coefficients, viscosity and thermal conductivity over wide density ranges. These data sets[5] and those from Adler et al. (1970)[28] will employed as our main data sources for the transport properties. For diffusion coefficients, due to a special feature mentioned above, more data sets are employed: Alder et al. (1970)[28], Easteal et al. (1983)[30], Erpenbeck and Wood (1991)[31], and Sigurgeisson and Heyes (2003)[8].

First and foremost, we test the predictive equation, Eq.(38) or (43), for viscosity, where the thermodynamic free volume is given by Eq.(25). The CS EoS[23] can also be used for the same purpose. Figure 5 depicts the results. As expected, the equation does not provide good prediction for low density range, $\rho^* < 0.3$, but otherwise works nicely. Considering the predictive nature, the model is successful.

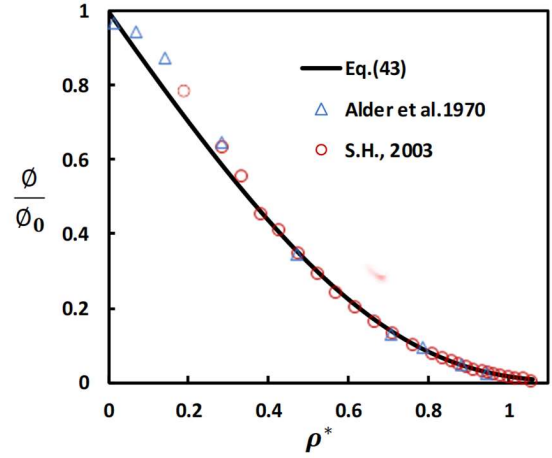

**Figure 5**. Fluidity $\emptyset = \mu^{-1}$ plot. The solid line is from Eq.(43) with Eq.(25) and $\xi = 0.75$. Data sources are ref[5,28].

For diffusion coefficient, the parameter in Eq.(44), $\chi = 0.836$, is from best fitting the MD data. Figure 6 presents the results. Again, in the intermediate range, $\rho^*$ from around 0.1 to 0.3, the power law shows slightly greater deviations. Hence, the new model is suggested for dense fluids or liquids, not for gases in the intermediate density range. By considering the facts that the HS diffusivity exhibits a maximum in the intermediate range and only one parameter is used in the model, the results are very satisfactory.

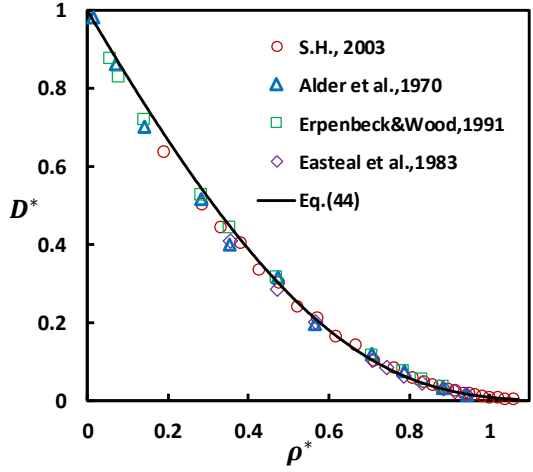

**Figure 6**. HS diffusion coefficient plot. Solid line is from eq.(44) with eq.(25); Data sources are ref[5,28,30,31].

For thermal conductivity, eq.(45) has been tested and the unique parameter, $\delta = 1.2$, was from fitting the MD data. As illustrated in Fig.7, the results are generally satisfactory, except at very high density region, the power law prediction gets worse.

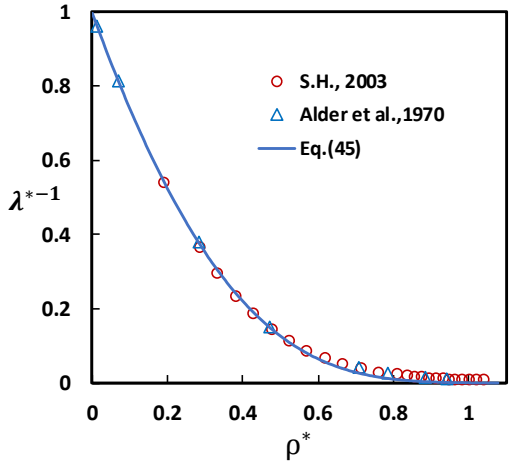

**Figure 7**. Thermal conductivity of HS fluid. Solid line: eq.(45). MD data sources are ref[5,28].

Finally, for demonstrating the applications of the hard sphere models to realistic fluids, here we extend the power law to Lenard-Jones fluid. For fluidity (viscosity), the equation for LJ fluid reads[8]:

$$\emptyset_{LJ}^* = \emptyset_{HS}^*(\sigma_{LJ}^e) exp\left(\frac{-\alpha_1}{T^*}\right) \quad (48)$$

where, $T^* = kT/\epsilon_{LJ}$, $\emptyset_{HS}^*$ is the expression for HS fluid, eq.(43). Due to the "softness" of the LJ fluid, in the HS model, the diameter is replaced by a so-called effective diameter for LJ fluid [8,10]:

$$\sigma_{LJ}^e = \sigma_{HS} f(T^*) \quad (49)$$

Generally, $\sigma_{LJ}^e$ could be density-dependent as well[10]. Some extensive studies have been carried for the LJ effective diameter[8-10]. In this work, we simply adopt the equation for $\sigma_{LJ}^e$ from previous works [8,10], namely, using the Boltzmann effective diameter:

$$\frac{\sigma_{LJ}^e}{\sigma_{HS}} = \alpha_0 \left[1 + \left(\frac{T^*}{T_0}\right)^{\frac{1}{2}}\right]^{-\frac{1}{6}} \quad (50)$$

where, $\alpha_0 = 2^{1/6}$, $T_0^{-1} = 1.3229$ [8].

No extra-parameter is required in the calculation of $\emptyset_{HS}^*(\sigma_{LJ}^e)$. The only parameter in the model is $\alpha_1$ (=0.1) introduced in the energy term, representing the contribution of attractive interactions. Figure 8 illustrates the results. In the wide temperature and density ranges, the predictions with only one parameter are very satisfactory. As reported by previous studies[8,9,33], over 95% of the property are from the hard sphere contribution, which confirms that the repulsive interaction dominates the properties of dense fluids, in particular, liquid.

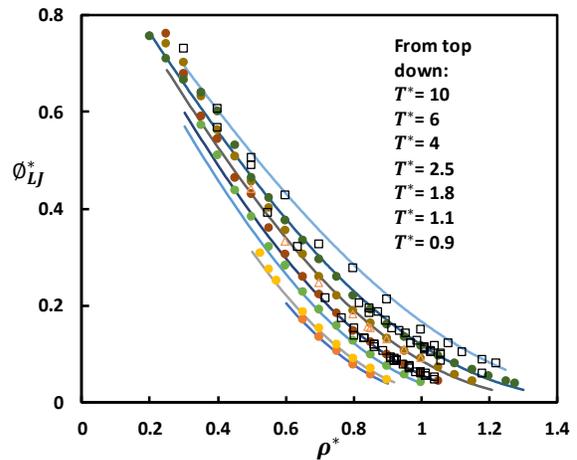

**Figure 8**. Plot of LJ fluid viscosity. Curves: prediction by eq.(48) with the energy term, $\alpha_1 = 0.1$. Circles: MD simulation from Meier et al. (2003)[32].

Similar equation can be written for diffusion coefficients of the LJ fluids:

$$D_{LJ}^* = D_{HS}^*(\sigma_{LJ}^e)\exp\left(\frac{-\alpha_2}{T^*}\right) \quad (51)$$

Where the same effective diameter given by eq.(49) and (50) is used, and the only parameter is $\alpha_2$ (= 0.1), fitted from MD simulation data [25]. Interestingly, the value of this parameter is the same as that in the viscosity model. The results are shown in Fig. 9. Similar to the viscosity case, the model shows greater deviations in the density range, $\rho^* = 0.1\ to\ 0.3$. In other words, all models should be applied to high density fluids or liquids.

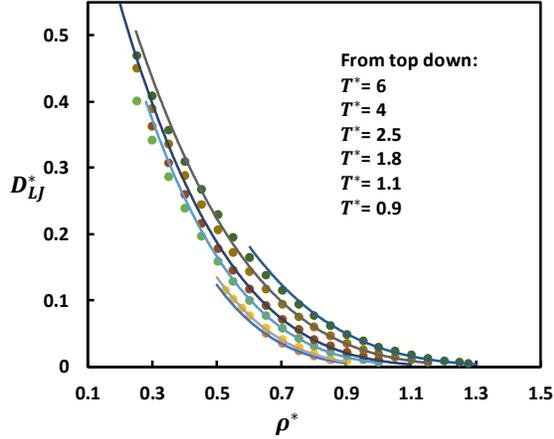

**Figure 9**. Diffusion coefficients for LJ fluid. Curves: calculated by eq.(51) with $\alpha_2 = 0.1$; circles: MD simulation from Meier et al. (2003) [32].

We also tried to apply the same approach for thermal conductivity of the LJ fluid, and it was found that the results are much worse compared to the cases of viscosity and diffusivity. This is not a surprise: thermal conductivity is not sensitive to particle diameter and therefore, the effective diameter method is not suitable for this property.

## 6. Discussions

There are numerous equations of state available for the HS fluid[34]. Here we are particularly interested in two EoS's, which are proposed by Heyes and Woodcock (1986)[35] and by Moshen-Nia et al. (1993)[36], respectively. The HW EoS[35] reads:

$$Z = 1 + \frac{4\eta}{(1-e\eta)^2} \quad (52)$$

from which thermodynamic free volume is given by:

$$\frac{v_{tf}}{v} = \exp\left(-\frac{e_1 v_0}{v - e_2 v_0}\right) \quad (53)$$

where $e = 1.175, e_1 = 2.96, e_2 = 0.87$. The MN EoS[36] reads:

$$Z = \frac{1 + c_2\eta}{1 - c_1\eta} \quad (54)$$

and thermodynamic free volume follows:

$$\frac{v_{tf}}{v} = (1 - c_1\eta)^{c_0} \quad (55)$$

where $c_0 = (c_1 + c_2)/c_1$; $c_2 = 2.48$, $c_1 = 1.88$, $c_0 = 2.32$.

Figure 10 depicts a comparison between the thermodynamic free volumes calculated from different HS EoS's. In addition, the MD results[2,14] for the geometric free volume are also plotted. First of all, we see (Figure 10b) that the change pattern of the geometric free volume aligns well with that from the thermodynamic free volume except at the lowest density point $\rho^* = 0.7$. This is comprehensible since at lower density region, MD simulation becomes less reliable. Secondly, the thermodynamic free volumes obtained from three EoS's are very close to each other, while in the high density (liquid) region there are visible discrepancies (with the results from Eq.(25) being most accurate). For practical applications, these discrepancies will not impact the final results. Lastly, as mentioned, the vdW free volume only works for low density gas. By the way, the match between the geometric free volume and the NM EoS, Eq.(55), is simply a coincident.

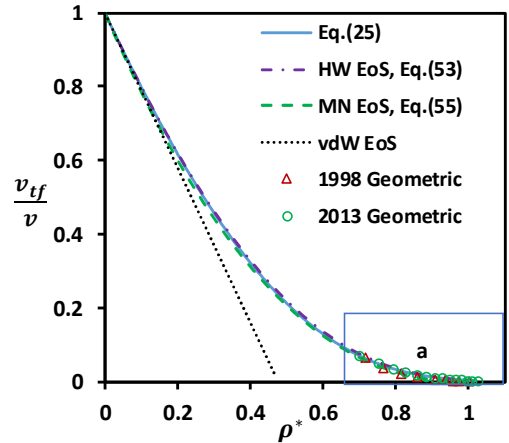

**Figure 10a**. Plots of thermodynamic and geometric free volumes over the entire density range.

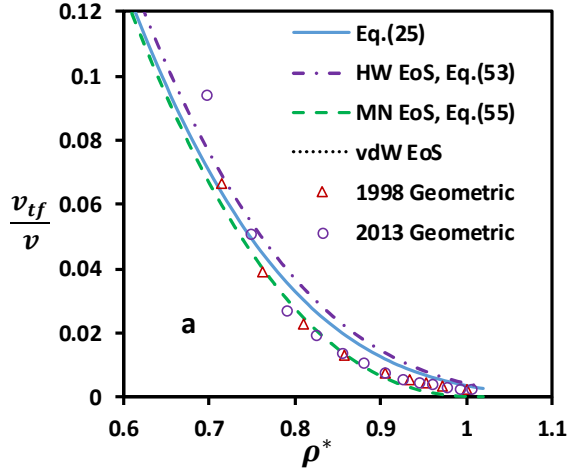

**Figure 10b.** Inset a of Figure 10a for high density region.

**Figure 10**. Comparison between different thermodynamic free volume expressions. Solid line: Eq.(25); dash-dotted line: Heyes and Woodcock 1986, eq.(53); dash line: Moshen-Nia et al. Eq.(55); dotted line, Van der Waals, eq.(4), $c = 4$. Geometric free volume data are from refs (2013)[2], (1998)[14].

An important conclusion drawn from the comparison is that both Eq.(53) and Eq.(55) can be used to calculate the thermodynamic free volume (hence the geometric free volume via Eq.(22)). Now we replace the thermodynamic free volume in Eq.(43) with the HW EoS, Eq.(53) and rearrange the constants:

$$\frac{\emptyset}{\emptyset_0} = \exp\left(-\frac{e_3 v_m}{v - v_m}\right) \quad (56)$$

where $e_3 = e_1 \xi / e_2$, $v_m = e_2 v_0$. Eq.(56) is a remarkable result: it is exactly the CTD model, eq.(9). Now we can explain the long-standing contradictory observations on the CTD model: the mathematical formula of the original Doolittle empirical equation[7], Eq. (3), is indeed "correct": it can be derived from the power law by utilizing Heyes and Woodcock EoS, eq.(53) for the thermodynamic free volume. The inappropriate expression for free volume, eq.(4), which works only for dilute gas, coincidently leads to the correct result. The general form, i.e. exponential law, is not suitable for dense fluids or liquids when a correct free volume expression is employed (Fig. 4).

The EoS proposed by Moshen-Nia et al. (1993)[36], Eq.(54) can also lead to an interesting result. A more recent EoS based on a geometric free volume theory has been proposed by Kegelberg et al. (2006)[15], which is similar to Eq.(54) with different value of $c_1$. By combining eq.(43) and Eq.(55), the power law can be re-written as:

$$\frac{\emptyset}{\emptyset_0} = (1 - c_1 \eta)^\tau \quad (57)$$

where $\tau = \xi c_0$. A similar equation can be derived for diffusion coefficient. This is another remarkable result. It's form coincides with a result from the mode coupling theory[4]. Sigurgeirsson and Heyes (2003) have empirically proposed the same equation for fitting their simulation data for HS fluid[5].

Lastly, the excess entropy is related to thermodynamic free volume by Eq.(21). Combining eq.(21) with, eq.(43), we have

$$\emptyset^* = \exp(-\lambda_0 s^{ex}) \quad (58)$$

Therefore, the scaling law, eq.(58), is only a different form of the power law as the relation between thermodynamic free volume and geometric free volume, eq.(22), is applied. With a similar treatment as viscosity, eq.(58), we can derive the following:

$$D^* = \exp(-\lambda_1 s^{ex}) \quad (59)$$

Eq.(58) and eq.(59) have been widely discussed in literature[3,29] as empirical laws. Here we decorate the scaling law with some physical justifications.

## 7. Conclusions

In this work a power-law relation between geometric free volume and transport properties is established for the HS fluid by using recent simulation results for the free volume distribution function. A correlation of the geometric free volume with the thermodynamic free volume (excess entropy) makes it possible to use well-developed EoS's to obtain various final expressions. It turns out that the final equations from several approaches can be reproduced by the power law. It is also shown that the most accepted exponential law is not suitable for the HS transport properties. In particular, the controversy regarding the well-know Cohen-Turnbull-Doolittle model is resolved by using the Heyes and Woodcock EoS.

The application of the HS model to real fluids is demonstrated by using the LJ fluids in which attractive interactions contribute a small percentage. For real fluids, especially liquids, the repulsive interaction still dominates the systems while the attractive interaction will play a bigger role[8,33].

## Acknowledgements

This work was started years ago and I had the privilege to discuss with Dr. Robin J. Speedy on free volume related topics. Dr. L. V. Woodcock has been very kind for proving answers to my questions. Those discussions helped a lot for me to start the work. I am also grateful to Dr. Krekelberg and Dr. Truskett for providing their MD simulation data.